\documentclass[%
 reprint,
 amsmath,amssymb,
 aps,
 pra,
]{revtex4-2}

\usepackage{graphicx}
\usepackage{dcolumn}
\usepackage{bm}
\usepackage{xcolor}
\usepackage{siunitx}
\usepackage{soul}
\DeclareSIUnit\gauss{G}
 \DeclareSIUnit\bohr{\text {\ensuremath {a}}_{0}}

\begin{document}    

\preprint{APS/123-QED}

\title{Fermion mediated pairing in the Ruderman-Kittel-Kasuya-Yosida to Efimov transition regime}

\author{Geyue Cai}
\thanks{These authors contributed equally to this work.}
\author{Henry Ando}
\thanks{These authors contributed equally to this work.}
\author{Sarah McCusker}
\author{Cheng Chin}
\affiliation{The James Franck Institute, Enrico Fermi Institute, and Department of Physics, \\ The University of Chicago, Chicago, IL 60637, USA}

\begin{abstract}
The Ruderman-Kittel-Kasuya-Yoshida (RKKY) interaction and Efimov physics are two distinct quantum phenomena in condensed matter and nuclear physics, respectively. The RKKY interaction describes correlations between impurities mediated by an electron gas, while Efimov physics describes universal bound states of three particles with resonant interactions. Recently, both effects have been observed in Bose-Fermi mixtures in the weak and resonant interaction regimes, respectively. Intriguing conjectures exist to elucidate how the two phenomena meet in the transition regime where the mixture is strongly interacting. In this work, we explore the RKKY-Efimov transition in a mixture of bosonic $^{133}$Cs and fermionic $^6$Li near a tunable interspecies Feshbach resonance. From dispersion and relaxation measurements, we find that the transition is highlighted by a fermion-mediated scattering resonance between Cs atoms and a weaker resonance on Li atoms. These resonances represent reactive scattering of Cs and Li atoms in the many-body regime, which reduces to an Efimov resonance in the thermal gas regime. Our observation demonstrates the intriguing interplay of two-, three-, and many-body physics in an Bose-Fermi mixture that connects condensed matter physics, nuclear physics and quantum many-body chemistry. 
\end{abstract}

\maketitle

Ultracold atoms offer a powerful platform to explore exotic quantum phases and quantum dynamics that are conjectured across physics disciplines ranging from condensed matter, to nuclear, to cosmological physics \cite{Bloch2012}. 
A unique advantage of cold atoms is their tunable interactions through Feshbach resonances \cite{Chin2010}, permitting precise and smooth tuning of the system from the weak to the strong interaction regime. Emergent phenomena, phase transitions, and novel quantum dynamics can thus be experimentally identified and compared with theoretical prediction in great detail \cite{Altman2021}.


\begin{figure}[ht!]
\center
\includegraphics[trim=0in 0in 0in 0in,clip,width=0.36\textwidth]{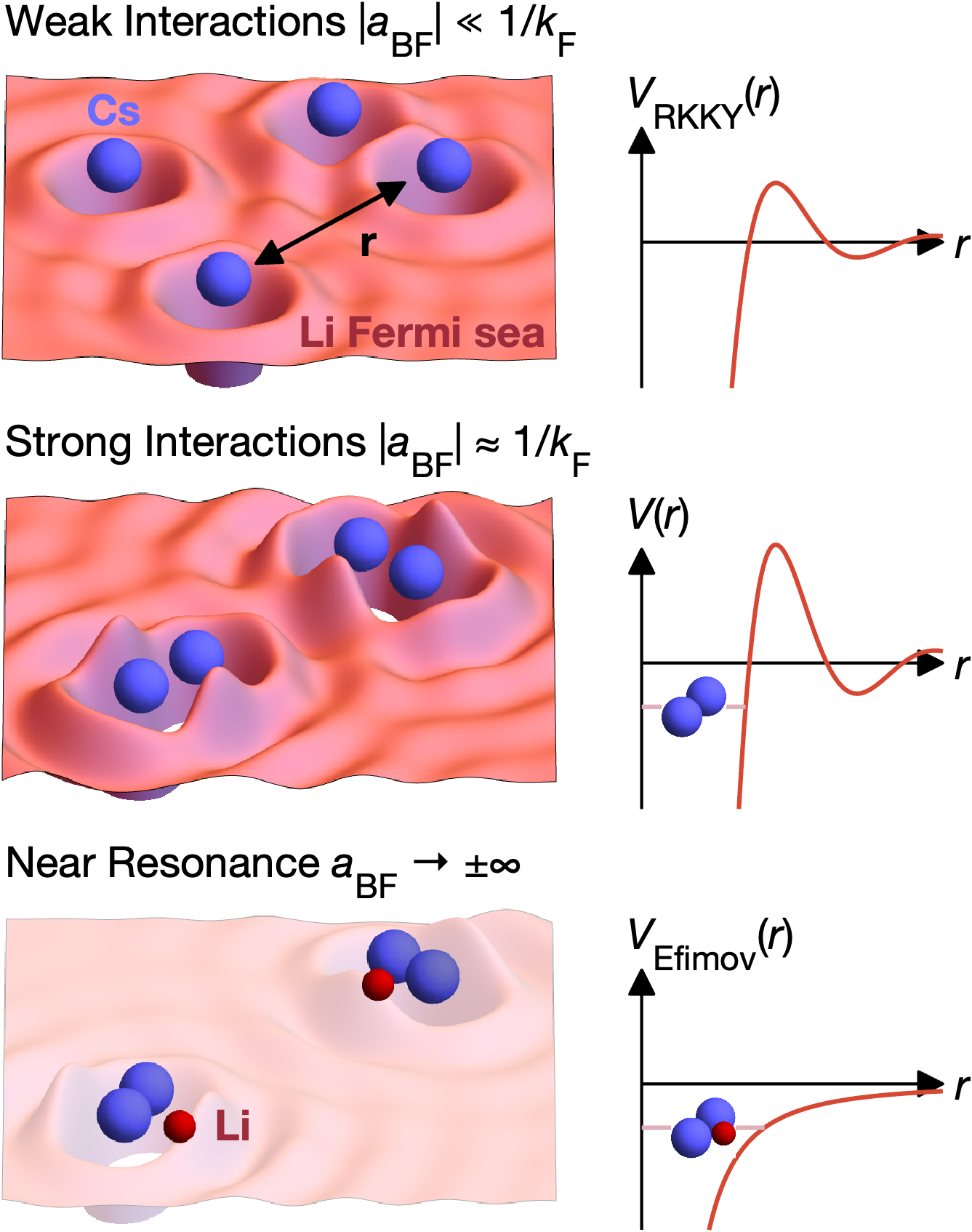}
\caption{\label{Fig1Illustration} Quantum mixtures of heavy bosonic $^{133}$Cs atoms (blue) and light fermionic $^{6}$Li atoms (red) in different interaction regimes. In the weak coupling regime with small interspecies scattering length $|a_\text{BF}|\ll 1/k_\text{F}$ (top panel), fermionic scattering leads to interactions between Cs atoms through the Ruderman–Kittel–Kasuya–Yosida (RKKY) mechanism. The RKKY potential $V_\text{RKKY}(r)$ shows Friedel oscillations due to fermion scattering (ripples in the Fermi sea) at the Fermi length scale $1/k_{F}$. In the strong coupling regime $|a_\text{BF}|\sim k_F^{-1}$ (middle panel), this paper reports binding of Cs atoms due to enhanced mediated interactions $V(r)$. Near the interspecies Feshbach resonance $a_\text{BF}\rightarrow \pm \infty$ (bottom panel), the effective potential supports Efimov $\text{Cs}_2\text{Li}$ states. Because of the large boson-fermion mass ratio, all effective potentials can be expressed as a function of the bosonic separation $r$.}
\end{figure}


Among the modern approaches to control atomic interactions, one promising development to engineer long-range potentials is based on mixtures of different atomic species. In cold atom mixtures, interactions between one species mediated by the other species have been demonstrated experimentally \cite{DeSalvo2019,Edri2020,Fritsche2021,Chen2022,Patel2023,Baroni2024}, and offer great flexibility to control the strength and the functional form of the effective potential \cite{Santamore2008,Nishida2009,De2014,AgruelloLuengo2022}. Such mediated interactions offer promising perspectives to simulate new classes of quantum systems in nature \cite{Paredes2024}, including long-range interactions between fundamental particles \cite{Weinberg1995}, superconducting materials \cite{Tinkham2004}, 
and magnetic orders in materials \cite{Ruderman1954, Kasuya1956, Yosida1957}.

Mixtures of particles with a large mass imbalance are particularly favorable for studying mediated interactions. The large mass ratio validates the Born-Oppenheimer approximation such that the light, fast moving particles induce an effective potential $V(r)$ between two  heavy, slowly moving particles separated by a distance $r$. Light bosons and heavy fermions, for example, are best to simulate phonon-mediated Cooper pairing in superconductors \cite{ Bijlsma2000,Heiselberg2000,Viverit2002a,Matera2003,Kinnunen2018, Efremov2002}. Halo trimer states were recently observed in a $^{23}$Na-$^{40}$K mixture \cite{Chuang2024}. Meanwhile, a heavy-boson light-fermion mixture can enhance the fermion-mediated interaction. This leads to a strong Efimov potential for boson-boson-fermion three-body systems \cite{Efimov1970,Naidon2017} and RKKY interactions for bosonic impurities embedded in a degenerate Fermi gas \cite{De2014}.

The mixture of bosonic $^{133}$Cs and fermionic $^{6}$Li atoms is an excellent platform to probe the above physics. The combination has both a large mass ratio of $m_{\text{Cs}}/m_{\text{Li}}\approx 22.1$ and convenient Feshbach tuning of the interspecies scattering length $a_\text{BF}$ \cite{Tung2013, Repp2013}. Because of the mass ratio enhancement, multiple Efimov states are observed near Feshbach resonances $a_\text{BF}\rightarrow\pm\infty$ \cite{Johansen2017,Pires2014}. In the weak coupling regime with $|a_\text{BF}|\ll k_F^{-1}$, where $k_F$ is the Fermi wavenumber, RKKY-type fermion-mediated interactions between Cs atoms are observed \cite{DeSalvo2019, Patel2023}. These works demonstrate the Efimov and RKKY effects in the resonant- and weak-interaction limit, respectively, see Fig.~1. 

In the transition regime where RKKY and Efimov physics meet, the Bose-Fermi mixture is strongly interacting with the scattering length comparable with the Fermi length scale $|a_\text{BF}|\approx k_F^{-1}$. 
Novel quantum phenomena are conjectured in this regime, including fermionic zero sound \cite{Shen2024}, $p$-wave fermionic superfluidity \cite{Nishida2009,Kinnunen2018}, Bose-Fermi droplets \cite{Rakshit2019}, and the decay of the mixture, which may simulate the collapse dynamics of white dwarf stars \cite{Karpiuk2021}. Relevant to this paper, the effect of a Fermi sea on few-body physics have been investigated theoretically \cite{Nygaard2014,Cui2014,Sun2019,Enss2020,Fisher2024,Guo2024}.

\begin{figure*}\centering
\includegraphics[trim=0in 0in 0in 0in,clip,width=\textwidth]{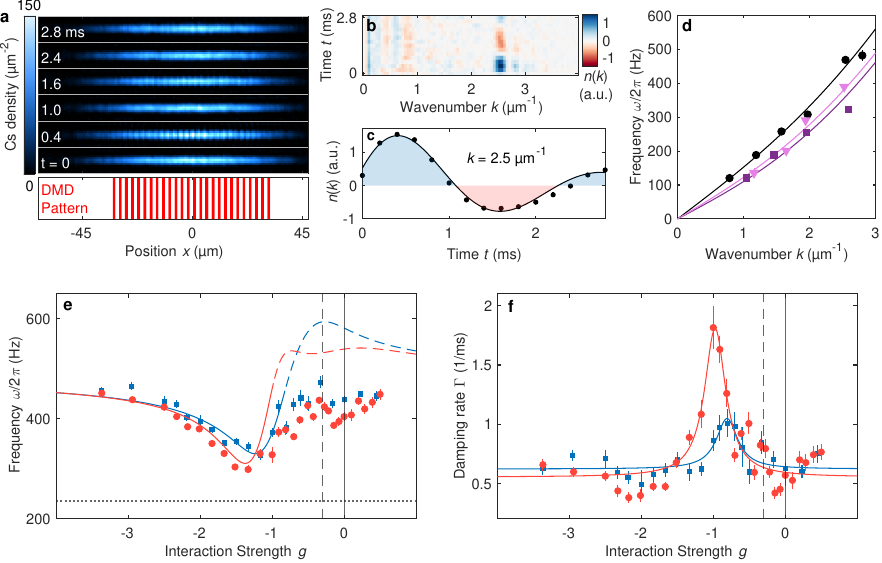}
\caption{\label{Fig2} Dispersion of Cs BEC embedded in a Li degenerate Fermi gas. (a) We imprint a periodic phase on the Cs condensate with a short $80~\mu$s pulse of a periodic optical potential, generated by a digital micromirror device (lower panel). The phase modulation at wavenumber $k=\SI{2.5}{\per\um}$ leads to an oscillating density wave order, shown in \textit{in situ} images of the BECs (upper panel) at an example interspecies scattering length $a_\text{BF}=-400~a_0$. (b) Fourier transform of the density profiles $n(k)$. (c) Real part of $n(k)$ is fit to $n = A e^{-\Gamma t}\sin \omega t$ (solid line), from which we extract the damping $\Gamma$, and frequency $\omega$. (d) Measured frequencies $\omega$ for $a_\text{BF}=\SI{-400}{\bohr}$ (black circles), $\SI{-688}{\bohr}$ (purple squares), and $a_\text{BF}\approx\pm\infty$ (pink triangles). Solid lines are fits to the data based on the Bogoliubov model. (e) Excitation frequency $\omega$ and (f) damping $\Gamma$ at $k=\SI{2.5}{\per\micro\meter}$ show a resonant dispersive and absorptive behavior for samples with 20,000 (red circles) and 10,000 Li atoms (blue squares). Fits based on an empirical model (solid lines), yield resonance positions at $a_\text{BF}=\SI{-1030(20)}{\bohr}$ for 20,000 Li atoms and \SI{-1230(50)}{\bohr} for 10,000 Li atoms. Dashed colored lines in panel (e) indicate the extension of the fit prediction beyond the fitting region, see \cite{Supplement} for more data and analysis. In all panels, error bars are 1-$\sigma$ standard deviations of the mean. In panels (e-f), the black solid line and dashed line indicate the Feshbach resonance and Efimov resonance, respectively. In panel (e), the black dotted line indicates the Cs free particle energy at $k=\SI{2.5}{\per\micro\meter}$.}
\end{figure*}

\begin{figure*}\center
\includegraphics[trim=0in 0in 0in 0in,clip, 
width=0.98\textwidth]{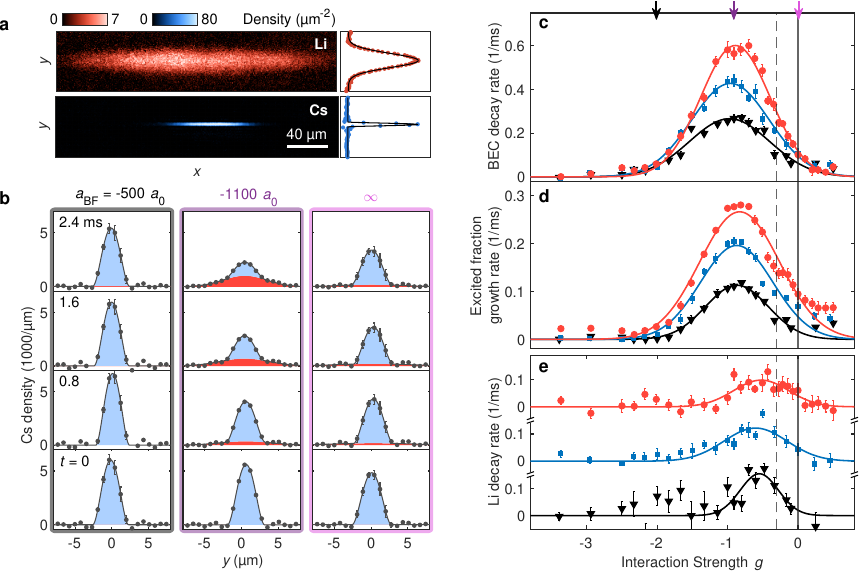}
\caption{\label{Fig3Resonance} Resonant decay and excitation Cs BECs induced by the Li degenerate Fermi gas. (a) Example in situ images of Li and Cs. Right panels show axially integrated density profiles. (b) Evolution of Cs density profile below (left), at (middle) and above (right) the mediated resonance at $a_\text{BF}=-1,100~a_0$. Bimodal fits to the density profile show the condensed (blue) and Gaussian (red) parts of the sample. The Gaussian fraction has an effective temperature of $\SIrange{130}{180}{\nano\kelvin}$. (c) Cs BEC number decay rate, (d) Cs excited fraction growth rate, and (e) Li number decay rate show resonances for samples with 20,000 (red circles), 10,000 (blue squares), and 5,000 (black triangles) 
Li atoms. Gaussian fits yield average resonance positions at $a_\text{BF}=\SI{-1090(50)}{\bohr}$ for the BEC decay, \SI{-1170(30)}{\bohr} for the excitation growth, and $\SI{-1800(100)}{\bohr}$ for the Li decay. In all panels, error bars are 1-$\sigma$ standard deviations of the mean. In panels (c-e), the Feshbach resonance (vertical solid line) and Efimov resonance (vertical dashed line) are shown for comparison.}
\end{figure*}

\begin{figure*}\center
\includegraphics[width=\textwidth]{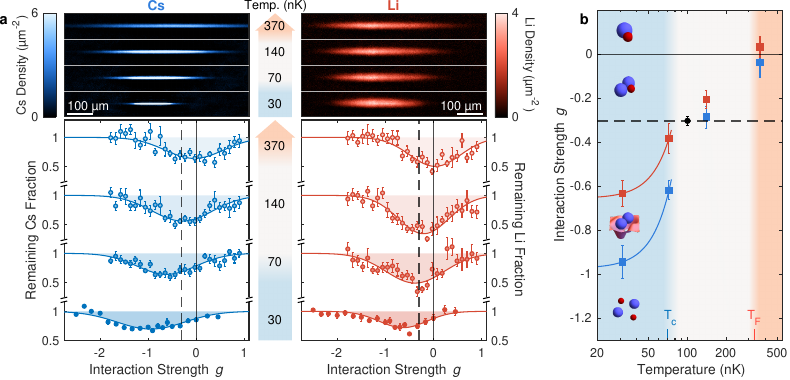}
\caption{\label{fig:figure4} Fermion mediated resonances in the thermal and quantum regimes. (a) Sample images with 30,000 Cs atoms and 20,000 Li atoms are prepared at 30, 70, 140, and \SI{370}{\nano\kelvin}. Lower panels show the Cs and Li surviving fraction after hold time $t_h = 2.8$, 3, 6, and \SI{10}{\milli\second} for the 30, 70, 140, and \SI{370}{\nano\kelvin} samples, respectively. We determine the positions of the loss peaks from Gaussian fits (solid lines). (b) Summary of loss centers in the classical and quantum regime. Here $T_c=75~$nK is the BEC critical temperature, and $T_\text{F}=360$~nK is the Fermi temperature. Li loss centers are shown in red, Cs in blue. The black circle is the measurement from Ref.~\cite{Johansen2017}. Red and blue solid lines are guides to the eye. Cartoons show the hypothesized many-body ground state: free Li (red balls) and Cs atoms (blue balls) for weak attractions $g<-0.9$, fermion-mediated Cs$_2$ pairs for $-0.9<g<-0.6$ and Cs$_2$Li trimers for $-0.6<g<0$. Finally CsLi dimers form above the Feshbach resonance $g>0$. In all plots, error bars are 1-$\sigma$ standard deviations of the mean. The solid and dashed lines indicate the Feshbach and the Efimov resonance, respectively. }
\end{figure*}

In this work, we investigate the Cs-Li quantum mixture in the ``RKKY-Efimov'' transition regime by tuning the interspecies interactions across a Feshbach resonance. We show that the transition physics is highlighted by a fermion mediated resonance between Cs atoms, when the the scattering length approaches the Fermi gas length scale $k_F|a_\text{BF}|\approx 1$. Resonant structure manifests in both the dynamical response of collective excitations as well as the decay and heating of the Cs condensate. The resonance position can be compared with theoretical model that incorporates both RKKY and Efimov physics. Our observations offer new insight into different fermion mediated two- and three-body binding processes in the many-body regime. 

Our experiments begin with a BEC of 30,000 $^{133}\text{Cs}$ atoms immersed in a single component degenerate Fermi gas of up to 20,000 $^{6}$Li atoms. The mixture is prepared in a single beam dipole trap with a weak axial ($x$) and tight transverse ($y$ and $z$) confinement. The trap frequencies are $\omega_{\text{Cs}} = (\omega_x, \omega_y, \omega_z)=2\pi \times (6.53, 153, 114)$~Hz for Cs and $\omega_{\text{Li}} =  2\pi \times (36,330,330)$~Hz for Li. Both species are spin-polarized in their lowest hyperfine ground state, where the interspecies scattering length $a_\text{BF}$ can be tuned near a Feshbach resonance at $\SI{892.65}{\gauss}$ \cite{Tung2013, Johansen2017}. Over this range, a bare Cs BEC is stable with a moderate boson-boson scattering length  $a_{\text{BB}}=270~a_0$ \cite{Berninger2013}, where $a_0$ is the Bohr radius. A bare Fermi gas is also stable and non-interacting. The mixture is initially prepared with weak attractive interactions at scattering length $a_\text{BF}=-300~a_0$ and a temperature of around \SI{30}{\nano\kelvin}. The BEC chemical potential is $\mu_\text{B} = k_{\text{B}}\times \SI{35}{\nano\kelvin}$, where $k_\text{B}$ is the Boltzmann constant, and the Fermi energy is up to $E_\text{F}=\hbar^2k_\text{F}^2/2m=k_{\text{B}}\times\SI{370}{\nano\kelvin}$, where $2\pi\hbar$ is the Planck constant and $m$ is the fermion mass. See Ref.~\cite{Patel2023} for details in the sample preparation.

To probe the mixture in the transition regime, we quickly switch the magnetic field near the Feshbach resonance \cite{Supplement} to vary the interspecies scattering length from weak attraction. We introduce an interaction strength $g=\SI{1000}{\bohr}/a_\text{BF}$ to interpolate the weak coupling RKKY regime $g\ll -1$ and the resonant Efimov regime $g\approx0$. The RKKY-Efimov transition occurs in our system with $-1<g\leq0$, where the mixtures become strongly interacting.

Our first experiment to probe the transition regime is to study the dispersion of the Cs BEC immersed in a Fermi gas. Soon after the magnetic field settles, we imprint a weak periodic phase pattern at spatial frequency $k$ on the sample, leading to the emergence of oscillatory density wave, see Fig.~2. From the \textit{in-situ} images on the Cs BEC, see Figs.~2b and c, we obtain the real and imaginary part of the dispersion $\omega(k)$ from the frequency $\omega$ and damping $\Gamma$ of the density wave response \cite{Shamass2012}. We perform the experiment for different $k$, scattering lengths $a_\text{BF}$, and Li atom numbers. 

In the weak interaction regime $|a_\text{BF}|< \SI{500}{\bohr}$, the density wave oscillates with low damping. The measured dispersion $\omega(k)$ can be described by the Bogoliubov dispersion of the BEC with an effective bosonic interactions $a_\text{eff}$ \cite{DeSalvo2019,Patel2023}, see Fig.~2d. The extracted $a_\text{eff}$ is consistent with the RKKY prediction \cite{DeSalvo2019} and sound speed measurements \cite{Patel2023}, see Fig.~2e. 

For greater attractions $|a_\text{BF}|>\SI{500}{\bohr}$, damping becomes significant. The extracted frequencies and damping rates can be described by the dispersive and absorptive response of a resonance: the frequency first reaches a minimum and then returns to the background value after passing a resonance, where the damping peaks, see Fig.~2f. Near and beyond the Feshbach resonance, the measured frequency deviates from our dispersive model, which we attribute to collision loss and the phase separation in the presence of interspecies repulsion \cite{Patel2023}. When we reduce the fermion number, the resonance strength weakens.  

To further understand the nature of the resonance, we investigate closely the decay and heating dynamics of Cs BECs embedded in the Fermi gas. Immediately after the magnetic field quench, we monitor the BEC and thermal fractions from the transverse density distribution, see Figs.~3a and 3b. In the weak coupling regime $|a_\text{BF}|<\SI{500}{\bohr}$, the BEC remains stable. Near the mediated resonance, however, fast decay of the BEC occurs within milliseconds, accompanied by fast generation of excited atoms. Beyond the mediated resonance, the decay and heating quickly reduce. No discernible resonance features are observed at the positions of Efimov and Feshbach resonances. The fitted resonance positions from Figs.~3c and 3d are consistent with the collective excitation measurement in Fig.~2. The weighted average of the resonance positions from these measurements is $a_1=\SI{-1100(100)}{\bohr}$ \cite{Supplement}. The resonance is absent without the degenerate Fermi gas.

We compare the measured resonance position with theoretical models. A calculation based on the RRKY potential shows that the mediated interactions can support a bound state of two impurities at \cite{Supplement} 
\begin{equation}
    a_\text{th} = \frac{\alpha}{k_F^*} \sqrt\frac{m}{M} = \SI{-1440}{\bohr},
\end{equation}
where $m$ is the mass of the light fermion, $M$ is the mass of the heavy impurity, $\alpha=-2.40...$ is a constant, and the wavenumber of the Fermi gas $k_\text{F}^*=\eta k_\text{F}$ is greater than that of an ideal Fermi gas $k_\text{F}$ because of the interspecies attraction. 

We estimate the enhancement factor $\eta=(1-g_\text{BF}n_\text{B}/E_\text{F})^{1/2}\approx 2.2$ from the mean-field model, consistent with our independent measurement $\eta=2.0\sim2.1$ \cite{Supplement}. Here $n_\text{B}$ is the boson density, $g_\text{BF} = 2\pi\hbar^2 a_\text{BF}/m_r$, and $m_r=Mm/(M+m)$ is the reduced mass of the impurities and fermion. An alternative theory that incorporates both the RKKY and the Efimov potentials in the strong coupling regime \cite{Enss2020} gives $a_\text{th}=\SI{-1600}{\bohr}$ \cite{Supplement}. Both predictions are in fair agreement with our measurement. 

The Li Fermi gas behaves differently from the Cs BEC. Near the mediated resonance, Li trap loss is much slower than the BEC damping and decay rates. This supports the picture that Li atoms only ``catalyze'' the collision resonance between Cs atoms. Pushing to even stronger interspecies attraction, a distinct loss peak appears in the Fermi gas at $a_2=\SI{-1800(100)}{\bohr}$. The resonance is persistent for different fermion densities, and is clearly different from the Efimov resonance at $a_\text{BF}=\SI{-3300(240)}{\bohr}$ and the Feshbach resonance at $a_\text{BF}=\pm\infty$, measured preciously in thermal mixtures of Cs and Li atoms \cite{Johansen2017}. 

What is the nature of the two resonances at $a_\text{BF}=a_1$ and $a_2$ in the transition regime, and how are they related to RKKY and Efimov potential? To elucidate the underlying processes, we monitor the resonant features as the mixture is prepared in the classical and quantum regime. Starting with a mixture at the lowest temperature $T\approx\SI{30}{\nano\kelvin}$, we quickly step up the optical trap intensity to excite the mixture and wait for the mixture to reach a new equilibrium. This process provides precise heating of the sample without discernible trap loss. We then perform trap loss measurement across the resonances, see Fig.~4.

We observe drastic changes in the resonance positions when crossing from the thermal regime into the quantum degeneracy regime. For thermal mixtures above the Cs BEC critical temperature $T>T_c$, both species show resonant features consistent with the Efimov resonance at $a_\text{BF}=\SI{-3300}{\bohr}$ and Feshbach resonance $a_\text{BF}=\pm\infty$, in agreement with previous works \cite{Johansen2017}. Below the BEC critical temperature $T<T_c$, the resonances appear at weaker interspecies attraction and a gap between the resonances in the Cs BEC and the Li Fermi gas widens. The result is summarized in Fig.~4b. 

Based on our observations, we propose a physical picture to elucidate the impact of quantum degeneracy to the Cs-Li mixtures in the RKKY-Efimov transition regime. As we increase the interspecies attraction from zero, the Fermi gas first mediates bosonic interactions through the RKKY mechanism. When the attraction strengthens and the scattering legnth reaches the critical value $a_\text{BF}=a_1=\SI{-1100(100)}{\bohr}$ or $g=-0.9(1)$ in our system, resonant scattering of bosons occurs, which we denote as a degenerate Fermi gas (DFG) mediated reaction process:

\begin{align}
  \text{Cs}+\text{Cs}\xrightarrow{\text{Li DFG}}\text{Cs}_2.
\end{align}

\noindent Enhancing the attraction further brings the system to the second threshold $a_2= \SI{-1800(100)}{\bohr}$ or $g=-0.55(4)$ in our system, where Li atoms are also bound to the boson pairs, denoted as another DFG mediated process:

\begin{align}
  \text{Li}+\text{Cs}_2\xrightarrow{\text{Li DFG}}\text{Cs}_2\text{Li}. 
\end{align}

\noindent Both processes in Eqs.~(2) and (3) are many-body in nature, discussed in Refs.~\cite{Fisher2024, Enss2020}. When the system enters the normal gas regime, the many-body processes merge into a single three-body process $\text{Cs}+\text{Cs}+\text{Li}\rightarrow\text{Cs}_2\text{Li}$, which is precisely the picture of an Efimov resonance \cite{Kraemer2006}. 

Our picture offers a consistent understanding of this and previous works \cite{DeSalvo2019,Patel2023} on interacting Cs-Li mixtures in all regimes. It also highlights the rich physics of fermion-mediated binding as a unique feature in the RKKY-Efimov transition regime. The observed smaller bosonic binding length $|a_1|$ and the larger fermionic binding length $|a_2|$ can be understood since heavy particles form a bound state more easily than light particles \cite{Supplement}. This is a trend that persists in the many-body regime, e.g., the mass scaling of Eq.~(1) favors pairing of heavier impurities.

We acknowledge T.~Enss, V.~Galitski, and K.~Wang for valuable discussions. We thank K.~Patel and M.~Rautenberg for valuable assistance with the experiment. This work is supported by the National Science Foundation under Grant No. PHY-2409612 and by the Air Force Office of Scientific Research under Award No. FA9550-21-1-0447. 
H.A. acknowledges support by the National Science Foundation Graduate Research Fellowship under Grant No. DGE 1746045. S.M. acknowledges support by the National Science Foundation under PHY-GRS 2103542. 

\bibliography{apssamp.bib}
\bibliographystyle{apsrev4-1}

\clearpage
 \renewcommand\thefigure{S\arabic{figure}}

\setcounter{figure}{0}   

\maketitle

\onecolumngrid
\begin{center}

\begin{large}
\textbf{Supplementary Material for\\Fermion mediated pairing of bosons in the strong coupling regime}
\end{large}\\
Geyue Cai, Henry Ando, Sarah McCusker, and Cheng Chin\\
\textit{ The James Franck Institute, Enrico Fermi Institute and Department of Physics,\\ The University of Chicago, Chicago, IL 60637, USA}
\vspace{1cm}
\twocolumngrid
\end{center}

\begin{table*}[!hbt]
\begin{tabular}{| c| c c c| c |}
\hline 
Measurement & 20,000 Li & 10,000 Li & 5,000 Li & Average \\
\hline
Dispersion $(a_0)$ & -1,030(20) & -1,230(50) & -1,220(90) &  -1,100(100)  \\
BEC decay rate $(a_0)$ & -1,120(10) & -1,040(20) & -1,030(30) &  -1,090(50)  \\
Excited fraction growth rate $(a_0)$ & -1,210(30) & -1,150(40) & -1,150(30) & -1,170(30)  \\
\hline
Li decay rate $(a_0)$ & -1,900(200) & -1,700(200) & -1,800(200) & -1,800(100)  \\
\hline
\end{tabular}
\caption{\label{tab:resonances} Table of resonance positions extracted from different quantities and different Li numbers in units of $a_0$. The Average column denotes a weighted average with error bars from the standard deviation. Centers are extracted from fits in Fig.~2 and 3. The additional 5,000 Li dataset in Fig.~\ref{sfig_fig2extra} is also included. The weighted average and standard deviation of all results for dispersion, BEC decay rate, and excited fraction growth rate are summarized as $a_1= \SI{1100(100)}{\bohr}$. The weighted average of the Li decay rate resonance positions are summarized as $a_2 = \SI{-1800(100)}{\bohr}$.}
\end{table*}

\begin{figure*}[!hbt]
\center
\includegraphics[trim=0in 0in 0in 0in,clip,width=\textwidth]{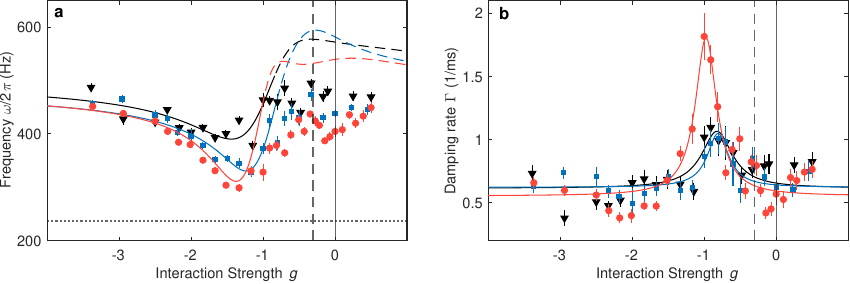}
\caption{\label{sfig_fig2extra} Dispersion of a Cs BEC embedded in a Li Fermi gas, with an additional dataset relative to Fig.~2. (e) Excitation frequency $\omega$ and (f) damping $\Gamma$ at the modulation wavenumber $k = \SI{2.5}{\per\micro\meter}$ show a resonant dispersive and absorptive behavior respectively, for samples with 20,000 (red circles) and 10,000 (blue squares) and 5,000 (black triangles) Li atoms. Fits based on an empirical model (solid lines), yield resonance positions at $a_\text{BF}=-1,030(20)$, $1,230(50)$, and $1,220(90)~a_0$ for 20,000, 10,000 and 5,000 Li atoms, respectively. Dashed colored lines in panel (e) indicate the extension of the fit prediction beyond the fitting region. 
In all panels, error bars are 1-$\sigma$ standard deviations of the mean. In both panels the vertical black solid line indicates the Feshbach resonance position, and the black dashed line indicates the Efimov resonance position. In panel (a), the black dotted line indicates the Cs free particle energy at wavenumber $k=\SI{2.5}{\per\um}$.}
\end{figure*}

\begin{figure}\center
\includegraphics[trim=0in 0in 0in 0in,clip,width=0.45\textwidth]{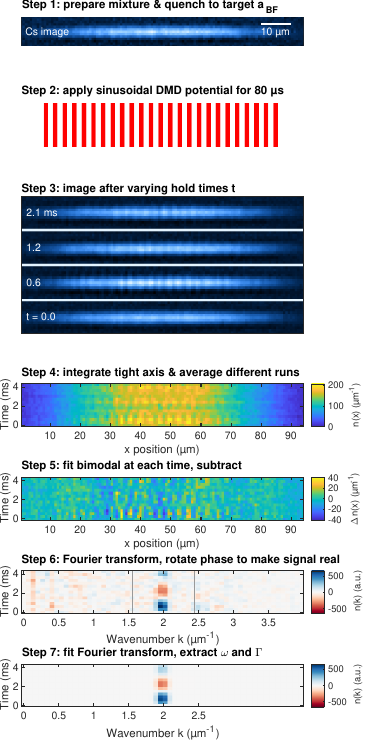}
\caption{\label{sfig:analysisdemo} Our analysis procedure for phase imprinting experiments. In Step 1, we prepare a degenerate $^{133}$Cs-$^{6}$Li Bose-Fermi mixture at weak attractive interactions ($a_\text{BF} = \SI{-300}{\bohr}$), before quenching in \SI{0.8}{\milli\second} to a target interactions strength. We then apply an \SI{80}{\micro\second} repulsive DMD potential with spatial frequency $k$ to the sample. We show here an example based on $k=\SI{2}{\per\micro\meter}$. In Step 3, we image after hold time $t_h$. In Step 4, we integrate along the radial direction and average the three rounds of data together to get the 1D axial density $n(x)$. In Step 5, we perform a bimodal fit to the density distribution and subtract off the fit to get a background-free view of the density waves, $\Delta n(x)$. In Step 6, we Fourier transform $\Delta n(x)$, then rotate the phase of the resulting Fourier transform to make the density wave signal real. We call the Fourier transform $n(k)$. The vertical gray lines in Step 6 show the extent of the fit region that we use in Step 7, where we perform a fit as described in Eq.~(\ref{eq:dampedfit}) to extract the frequency $\omega$ and damping $\Gamma$ of the density wave oscillations. 
}
\end{figure}

\subsection{\textit{In situ} imaging of Li and Cs atoms}\label{sec:imaging}


We implement quantum gas microscopy on both Li and Cs atoms based on absorption imaging. We use the ``fast kinetics mode'' of an Andor Ikon-M CCD camera to take single-shot images of both atomic species in one experiment. Imaging resolution on both species is \SI{1.5}{\micro\meter}. When imaging Li first, we expose the top 140 rows of the CCD pixels to the 671~nm imaging beam in the presence of the Li atoms, then shift down the stored charges into the dark region before exposing the 852~nm imaging beam in the presence the Cs atoms. This process continues for the bare 671~nm imaging beam and the bare 852~nm imaging beam to record background light level. The minimal shift time, and thus the time between imaging the two species, is \SI{0.42}{\milli\second}. This is a short enough time that the heavy Cs atoms should move no more than 1 pixel or \SI{0.78}{\micro\meter}. However, we find that taking Li images first can reduce the data quality of the phase imprinting experiments, so for Figs.~2 and 3 we image Cs before Li.

To calibrate the \textit{in situ} density measurement, we implement the absorption imaging calibration \cite{Veyron2022} for our Cs samples. This calibration constrains systematics due to radiative reabsorption in a density sample. This method improves the agreement between the measured density profiles of bare Cs atoms and the expectation based on the Thomas-Fermi approximation.

\subsection{Phase imprinting experiment}\label{sec:phaseimprint}

For the data shown in Fig.~2, our experimental procedure is shown in Fig.~\ref{sfig:analysisdemo}. Starting with a degenerate mixture prepared at interspecies scattering length $a_\text{BF} = -300~a_0$, we apply a carefully engineered field jump sequence which reaches the target magnetic field within \SI{0.8}{\milli\second}, after which the field is stable within \SI{10}{\milli\gauss}. The beginning of the field jump is defined as time $t=\SI{-1}{\milli\second}$.

We then apply an \SI{80}{\micro\second} repulsive DMD pulse at $t = \SI{-0.08}{\milli\second}$, \SI{0.92}{\milli\second} after initiating the field jump. Details of our DMD projection system can be found in Ref.~\cite{Patel2023}. For the main data in Figs.~2e-f, the DMD pattern is a simple 100$\%$-on 100$\%$-off vertical stripe pattern. For longer wavelength patterns such as those depicted in Fig.~2d, we use a pattern that approximates a sinusoidal intensity using the error diffusion halftoning algorithm, as the simple binary pattern produces noticeable higher spatial harmonics. By varying the DMD pulse duration, we verified that the measured density wave amplitude is linear in the DMD pulse length, showing that we are not exciting a small fraction of the condensed atoms into the $\pm k$ modes. 

After applying the DMD pulse, we wait a varying hold time $t_h$ before performing \textit{in situ} imaging on Cs atoms. We perform three rounds of the experiment at each scattering length and each value of $t_h$ for averaging. The raw images are shown in Figs.~2 a and 3 a.

To measure the dispersion as shown in Fig.~2 d, we integrate the images along the tight axis to produce 1D density profiles. We then subtract off a bimodal background distribution to produce a background-free view of the density wave. Next, we Fourier transform the data, and adjust the phase of the resulting Fourier spectrum $n(k,t)$ such that the datum with the largest amplitude (typically at the wavenumber of our applied modulation) is real and positive for small hold time $t_h$.

Finally, we perform a 2D fit to the real part of this resulting Fourier transform. The fit function includes a Gaussian at wavenumber $k=k_0$ and an underdamped oscillation in the hold time, 
\begin{equation}\label{eq:dampedfit}
    n_\text{fit}(k,t) = A e^{-(k-k_0)^2/2\sigma_k^2}e^{-\Gamma t}\sin{\omega t},
\end{equation}
where $A, k_0, \sigma_k, \Gamma$, and $\omega$ are fit parameters, and we only use the data from a small region around $k$ for the fit. This model fits the data with phase imprinting wavenumber $k=\SI{2.5}{\per\micro\meter}$ reasonably well, as the density wave appears underdamped in all but the most rapidly decaying condensates (around $a_\text{BF} \approx \SI{-1100}{\bohr}$).

In Fig.~2d, we adopt a fit based on the Bogoliubov dispersion 
\begin{equation}\label{barebogliubov}
    \omega(k) =  
    \frac{1}{\hbar}\sqrt{\epsilon_k^2 + 2 \beta n_\text{B} a_\text{eff} \epsilon_k }
\end{equation}
to the measured $\omega(k)$, where $n_\text{B}$ is the average boson density across the condensate, $\epsilon_k=\hbar^2k^2/2M$ is the free particle dispersion, $\beta=4\pi \hbar^2/M $, $M$ is the mass of a Cs atom, and $a_\text{eff}$ represents the effective scattering length between bosons \cite{Pethick2002,Kavoulakis1998}. Theoretical calculations on the dispersion for an interacting Bose-Fermi mixture are given in Refs.~\cite{Yip2001,Zheng2021,Viverit2002,Zheng2024}.

In Figs.~2e-f, we employ an empirical model and a multi-step fitting procedure to fit the data and extract the resonance position. The data in Figs.~2e-f are reproduced in Fig.~\ref{sfig_fig2extra}. We assume that the effective boson-boson scattering length $\tilde{a}_\text{eff}$ may be complex, and takes the form 
\begin{equation}
    \tilde{a}_\text{eff} = a_\text{BB} + 
    \frac{\Delta}{1/a_\text{BF}  - 1/a_\text{res} + i \gamma},
\end{equation}
where $a_\text{BF}$ is the boson-fermion scattering length, $a_\text{BB}$ is the known background boson-boson scattering length, and $\Delta$, $a_\text{res}$, and $\gamma$ are the fitting parameters corresponding to the width, central position, and decay strength of the resonance. 

Plugging this model into the Bogliubov dispersion Eq.~(\ref{barebogliubov}) gives 
\begin{gather}
    \omega + i\Gamma =
    \sqrt{\epsilon_k^2 +2 \beta n_\text{B} \tilde{a}_\text{eff} \epsilon_k  },\label{complexomegagamma}
\end{gather}
where $\Gamma$ is the quasiparticle decay rate. Due to the strong BEC decay we report in Fig.~3, the BEC density $n_\text{B}$ depends on the interspecies scattering length $a_\text{BF}$. We thus employ a Gaussian fit to the BEC loss to obtain the time-averaged BEC density for each $a_\text{BF}$, and apply the corrected Cs density in the fits. Nonetheless, a direct, simultaneous fit of Eq.~(\ref{complexomegagamma}) to the quasiparticle dispersion and damping does not match the data well near and above the mediated Feshbach resonance. 

We adopt a two-step fitting process for the data shown in Figs.~2e-f. First, we fit a Lorentzian to the damping data (Fig.~2f) to extract a resonance position. We then fit the frequency data (Fig.~2e) to the real part of Eq.~(\ref{complexomegagamma}). This model predicts a suppression of the excitation frequency $\omega$ below the resonance and an increase in $\omega$ to the right of resonance. Our datasets agree with the model below the resonance, but deviate from the model above the resonance, see Fig.~2e. We attribute the deviation to the fast three-body loss and phase separation, which is not properly accounted for in our model. Thus, for Fig.~2e we fit only the $\omega$ data to the left of the resonances. In Fig.~2e, we indicate the fit region by solid lines and the unfitted region with dashed lines. 

An additional dataset with 5,000 Li atoms is omitted from Fig.~2 for visual clarity. It is provided here in Fig.~\ref{sfig_fig2extra}. Resonance positions extracted from the fits are reported in Table~\ref{tab:resonances}, which also includes the measured resonance positions from Fig.~3.

\subsection{Preparation of Cs-Li samples at different temperatures}
In Fig.~4, we prepare mixtures at varying temperatures. For these experiments, we mostly follow the normal preparation and evaporation procedure of Figs.~2-3, but rather than holding the sample at the lowest trap depth for \SI{500}{\milli\second} before imaging, we hold the sample at the lowest trap depth for  only \SI{200}{\milli\second}, then heat the mixture by jumping the trap depth to a deeper value before waiting the remaining \SI{300}{\milli\second}. We find that this \SI{300}{\milli\second} hold provides enough time for the density profiles of both species to stop changing, and by staying at the deeper trap depth we retain similar atom numbers in both species as compared to the degenerate sample. 

We extract the temperature by performing a bimodal fit to the Cs axial density profile. Assuming the axial trap frequency (which mostly comes from magnetic trapping) is still $\omega_x=2\pi\times\SI{6.53}{\hertz}$, the temperature is related to the width $\sigma$ of the Gaussian part of the bimodal distribution through $k_B T = M\omega_x^2\sigma^2$. To fit the small thermal fraction at \SI{35}{\nano\kelvin}, we further constrain our fit by enforcing that the condensate fraction $N_0/N = 1-(T/T_c)^3$, where $T_c$ is the BEC critical temperature.

Our Fig.~4 analysis procedure is straightforward: we take dual-species \textit{in situ} images at each $a_\text{BF}$ value and each temperature, fit a bimodal (Gaussian) distribution to the Cs (Li) radial density profile, and extract atom numbers from each. We perform two calibrations as we go. Firstly, after each full data round (each $a_\text{BF}$ value and each temperature), we perform a magnetic field calibration to determine the electromagnet current corresponding to the magnetic field of the Feshbach resonance. We then interpolate this current-field calibration across the intervening data to get an accurate magnetic field value for each datum, and then bin the data before averaging to get Fig.~4. Secondly, rather than assuming a constant initial Cs and Li number, we take data at both $t=0$ as well as $t=t_h$, then compensate for systematic differences in initial number by reporting the final / initial number instead of just final number.


\subsection{Li Density Enhancement}

To estimate the local enhancement of $k_\text{F}$ due to the interspecies attraction, we work within the Thomas-Fermi approximation. Additionally, since the Fermi gas's dynamical time scale (given by $\hbar/E_\text{F} \approx \SI{23}{\micro\second}$) is much shorter than the time scale of the magnetic field jump (\SI{0.8}{\milli\second}), we expect the Fermi gas is able to follow the magnetic field jump. Thus, after jumping to the new interaction strength $g_\text{BF}$, we have 
\begin{equation}
    E_\text{F}= \frac{(\hbar k_F^*)^2}{2m_\text{F}} + g_\text{BF}n_\text{B} + V,
\end{equation}
where  $k^*_\text{F} = (6\pi^2n^*_\text{F})^\frac{1}{3}$ is the local fermion wavenumber, $n^*_\text{F}$ is the local fermion density modified by the interspecies interaction $g_\text{BF}$ and $V$ is the external trapping potential. The peak value of the enhanced local Fermi wavenumber $k^*_\text{F}$ is then
\begin{equation}
    k^*_\text{F} = \sqrt{\frac{2m_\text{F}}{\hbar^2}(E_\text{F}-g_\text{BF}n_\text{B})},
\end{equation}
giving an enhancement factor 
\begin{equation}
    \eta = \frac{k^*_\text{F}}{k_\text{F}} = \sqrt{1-g_\text{BF}n_\text{B}/E_\text{F}}.
\end{equation}
This predicts $\eta\approx2.2$ near the observed resonance at $a_1 = \SI{-1100}{\bohr}$.

An independent confirmation of the fermi gas density enhancement comes from \textit{in situ} images of Li degenerate Fermi gas. Enhanced Li density within the Cs BEC is observed with large interspecies attraction. A strong enhancement of the fermion density on the Cs BEC of $8\sim10$ for $-500~a_0<a_\text{BF}<\SI{-1000}{\bohr}$ is observed, which corresponds to $\eta=2.0\sim2.1$.


\begin{figure}\center
\includegraphics[trim=0in 0in 0in 0in,clip,width=0.45\textwidth]{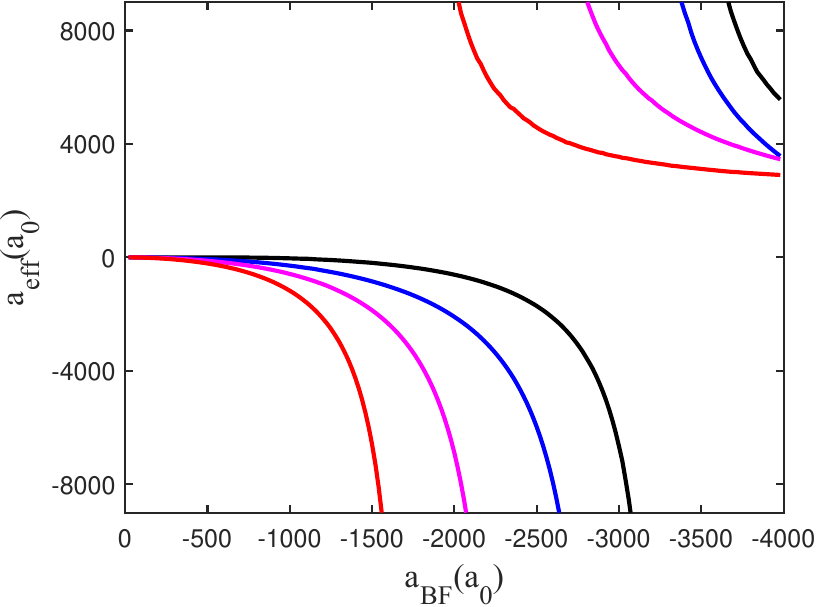}
\caption{\label{sfig:resPosition} 
Effective scattering length between Cs atoms in Li Fermi gas with density $\approx 0$ (black), 0.1~$n_\text{F}$ (blue), $n_\text{F}$ (magenta), and 10~$n_\text{F}$ (red). Here $n_\text{F}=4.7\times 10^{11}$cm$^{-3}$ is the nominal density of a non-interacting Li Fermi gas in our experiment while 10~$n_\text{F}$ is approximately the maximum fermion density on our Cs BEC in the presence of strong interspecies attraction. Divergence of the scattering length shows the pairing resonances shifting from the Efimov resonance at $a_\text{BF}= \SI{-3300}{}$, \SI{-2980}{}, \SI{-2360}{}, to \SI{-1720}{\bohr} with increasing densities. The calculation is based on the model in Ref.~\cite{Enss2020}.}
\end{figure}

\subsection{Resonance position of the fermion-mediated potential}
We present two estimations on the resonance position of fermion-mediated interactions in the main text. One is based on the leading order RKKY potential $V_\text{RKKY}(R)$ and the second model incorporates both RKKY and Efimov effects in the strong coupling regime, following the calculation in Ref.~\cite{Enss2020}, which builds on Ref.~\cite{Nishida2009}.

Consider two heavy impurities with mass $M$ embedded in a degenerate fermi gas of light atoms with mass $m\ll M$ and Fermi energy $E_F=\hbar^2k_\text{F}^2/2m$. Based on the Born-Oppenheimer approximation, the effective potential between the impurities separated by $R$ is given in the limit of weak interaction $\epsilon =|k_\text{F}a_\text{BF}| \ll 1$ by \cite{Tsurumi2000,Chui2004,Santamore2008,Nishida2009}

\begin{equation}
    V_{\text{RKKY}}(R) = E_{\text{F}}\frac{8\epsilon^2}{\pi}
    \frac{r  \cos r - \sin r }{ r^4} + O(\epsilon^3),
\end{equation}

\noindent where $r=2k_\text{F}R$ is the reduced separation. 

We describe the wavefunction of two impurities interacting with $V_\text{RKKY}$ with the Schroedinger's equation  

\begin{equation}
    \left[-\frac{\hbar^2\nabla^2}{M}+V_{\text{RKKY}}(R)\right]\psi(R)=E\psi(R), 
\end{equation}

\noindent which for strong enough attraction can support a bound state at $E=0$. The condition for the emergence of the bound state can be estimated by solving the variable phase equation for the scattering length function $a(R)$ \cite{Babikov1967}, which yields

\begin{equation}
    \frac{da(R)}{dR}=-\frac{M}{\hbar^2}V_\text{RKKY}(R)[R-a(R)]^2.
\end{equation}

\noindent We can then calculate the scattering length $a_\text{eff}=\lim_{R\rightarrow\infty}a(R)$, and the bound state emerges when the scattering length diverges $a_\text{eff}\rightarrow -\infty$. 

Integrating the variable phase equation numerically with the RKKY potential up to leading order in $\epsilon$, we obtain the condition on the fermion mediated binding as 

\begin{equation}
    k_\text{F} a_\text{th}=-2.40...\sqrt{\frac{m}{M}}.
\end{equation}

Remarkably, in systems with a large mass ratio $M/m\gg 1$, impurity binding occurs in the perturbation regime with $|k_Fa_\text{th}|\ll 1$, which validates the calculation based on the leading order RKKY potential. Here $a_\text{th}$ should be compared with $a_1$ in our experiment. 

Our second estimation is based on a calculation that takes both the RKKY and the Efimov three-body effects into account in the strong coupling regime~\cite{Enss2020}. Consider two Cs atoms separated by $R$ immersed in a Fermi gas. The calculation on the effective potential $V(R)$ assumes a large mass ratio $M\gg m$ and reproduces the RKKY potential in the weak interspecies interaction limit, and the Efimov potential in the limit of dilute Fermi gas. Here we summarize our calculation based on the paper. We calculate the Cs-Cs scattering property from $V(R)$ with the variable phase equation, see Fig.~\ref{sfig:resPosition}. Pairing resonances occur when the scattering length flips sign. 


The theory suggests a clear shift of the resonance position toward smaller interspecies scattering length $a_\text{BF}$ when fermion density increases. This confirms a stronger mediated attractions in a denser Fermi gas and thus a smaller interspecies attraction can bind Cs atoms. In the opposite limit when the Fermi gas density approaches zero, we recover the Efimov resonance.

The theory also elucidates the difference between the boson binding length $a_1$ and fermion binding length $a_2$. Fermions are bound to boson pairs when the magnitude of the interspecies scattering length $|a_\text{BF}|$ exceeds the size of the pairs \cite{Enss2020}. As $k_\text{F}$ characterizes the length scale of the fermion mediated potential and thus the boson pair size, we obtain the condition to bind fermions as $|a_2|\approx k_{F}^{-1}$. Thus we conclude a much stronger attraction $|a_2|>|a_1|$ is needed to bind light fermions than heavy bosons, which is in agreement with our observation.  

\end{document}